# Atomic-Scale Imaging of Fractional Spinon Quasiparticles in Open-Shell Triangulene Spin-1/2 Chains

Zhangyu Yuan[1,#], Xin-Yu Zhang[3,#], Yashi Jiang[1,#], Xiangjian Qian[1], Ying Wang[3], Yufeng Liu[1], Liang Liu[1,4], Xiaoxue Liu[1,4], Dandan Guan[1,4], Yaoyi Li[1,4], Hao Zheng[1,4], Canhua Liu[1,4], Jinfeng Jia[1,4], Mingpu Qin[1,4,*], Pei-Nian Liu[2,3,*], Deng-Yuan Li[2,*], Shiyong Wang[1,4,*]

[1]Key Laboratory of Artificial Structures and Quantum Control (Ministry of Education), TD Lee Institute, School of Physics and Astronomy, Shanghai Jiao Tong University, 800 Dongchuan Road, Shanghai 200240, China

[2]Key Laboratory of Natural Medicines, Department of Medicinal Chemistry, China Pharmaceutical University, Nanjing, 211198, P. R. China

[3]Key Laboratory for Advanced Materials and Feringa Nobel Prize Scientist Joint Research Center, Frontiers Science Center for Materiobiology and Dynamic Chemistry, State Key Laboratory of Chemical Engineering, School of Chemistry and Molecular Engineering, East China University of Science & Technology, Shanghai, 200237, P. R. China

[4]Hefei National Laboratory, Hefei 230088, China

[#]These authors contributed equally.

*Corresponding Authors: qinmingpu@sjtu.edu.cn, liupn@cpu.edu.cn, dengyuanli@cpu.edu.cn, shiyong.wang@sjtu.edu.cn.

**Abstract**

The emergence of spinon quasiparticles, which carry spin but lack charge, is a hallmark of collective quantum phenomena in low-dimensional quantum spin systems. While the existence of spinons has been demonstrated through scattering spectroscopy in ensemble samples, real-space imaging of these quasiparticles within individual spin chains has remained elusive. In this study, we construct individual Heisenberg antiferromagnetic spin-1/2 chains using open-shell [2]triangulene molecules as building blocks. Each [2]triangulene unit, owing to its sublattice imbalance, hosts a net spin-1/2 in accordance with Lieb's theorem, and these spins are antiferromagnetically coupled within covalent chains with a coupling strength of $J = 45$ meV. Through scanning tunneling microscopy and spectroscopy, we probe the spin states, excitation gaps, and their spatial excitation weights within covalent spin chains of varying lengths with atomic precision. Our investigation reveals that the excitation gap decreases as the chain length increases, extrapolating to zero for long chains, consistent with Haldane's gapless prediction. Moreover, inelastic tunneling spectroscopy reveals an *m*-shaped energy dispersion characteristic of confined spinon quasiparticles in a one-dimensional quantum box. These findings establish a promising strategy for exploring the unique properties of excitation quasiparticles and their broad implications for quantum information.

**Introduction**

Understanding the behavior of quasiparticles—emergent entities that arise from the collective interactions of many particles—has become a central theme in quantum materials[1–3]. Among these, spinons are a particularly intriguing class of quasiparticles that emerge in low-dimensional quantum spin systems[4–10]. Unlike ordinary particles, spinons carry spin without an accompanying charge. This phenomenon is a direct manifestation of the complex quantum correlations presented in strongly interacting electron systems, especially in reduced dimensions where quantum fluctuations are enhanced. Spinons were first theorized in the context of 1D quantum antiferromagnets, where a spin-1 excitation splits into two independent spin-1/2 excitations, known as spinons[4,11–13]. These quasiparticles have been indirectly observed through spectroscopic techniques, such as neutron scattering and resonant inelastic X-ray scattering, in bulk materials that exhibit 1D spin chain behavior[12,14]. However, direct real-space observation of spinons in individual spin chains remains a significant challenge, primarily due to the difficulty in isolating and probing these systems with sufficient precision.

Open-shell molecules, especially those derived from polycyclic aromatic hydrocarbons, have emerged as promising candidates for constructing and studying quantum magnetism at the molecular level[15–32]. Among these, triangulene—a triangular benzenoid hydrocarbon with unpaired electrons—has garnered significant attention[17,33–36]. Due to its sublattice imbalance, triangulene exhibits a tunable net spin, making it an ideal building block for creating artificial quantum spin systems[37]. When triangulene units are linked together, they form quantum spin systems with tunable exchange interactions, offering a molecular platform for investigating many-body quantum properties[38–42]. Leveraging advanced techniques such as scanning tunneling microscopy and spectroscopy (STM/STS), previous studies have achieved atomic-level precision in probing spin states and excitations within open-shell molecular systems[16,20,23,43–53]. The interplay between the intrinsic spin properties of open-shell molecules and their collective behavior in spin chains presents a unique opportunity to explore spinon quasiparticles in a controlled and tunable environment, a feat that has yet to be realized.

In this study, we leverage the unique properties of open-shell [2]triangulene to construct individual Heisenberg antiferromagnetic spin-1/2 chains on Au(111) by on-surface synthesis. Utilizing STM/STS, we probe the spin states, excitation gaps, and their spatial excitation weights within these chains, with a focus on understanding how these properties evolve with chain length. Our findings reveal that the excitation gap decreases as the chain length increases, approaching zero for an infinitely long chain—a behavior consistent with theoretical predictions for 1D quantum spin-1/2 antiferromagnets. Additionally, inelastic tunneling spectroscopy uncovers an *m*-shaped energy dispersion indicative of confined spinon quasiparticles, which we successfully model using density matrix renormalization group (DMRG) and exact diagonalization calculations. This work not only provides direct insight into the behavior of fractional spinon quasiparticles in a controlled molecular system but also highlights the potential of open-shell triangulene-based spin chains as a versatile platform for exploring fundamental aspects of quantum magnetism. The ability to manipulate and observe spinons in engineered molecular systems may also pave the way for advances in quantum computing, spintronics, and other quantum-based applications.

**Results and discussions**

**On-surface synthesis of closed-shell phenalene chains.**

Open-shell [2]triangulene is highly reactive and challenging to obtain through conventional wet chemistry. Instead, we use closed-shell $Br_2$-phenalene as a precursor for the on-surface synthesis of phenalene chains on Au(111)[54,55]. As shown in Fig. 1a, thermal annealing of $Br_2$-phenalene precursors on Au(111) simultaneously activates both debrominative and dehydrogenative C-C coupling, resulting in the formation of unexpected oligomers randomly fused by five-membered rings (Extended Fig. E1). To selectively facilitate debrominative C-C coupling while preventing dehydrogenative C-C coupling, we utilize alkali potassium metal atoms as catalysts which are expected to decrease the debromination temperature[56]. Mild annealing at approximately 338 K in the presence of potassium (K) adatoms produces covalent phenalene oligomers of varying lengths (Fig. 1b). We have performed density functional theory (DFT) calculations to obtain the reaction barriers with and without K catalysts, where K adatom reduces the barrier of debromination (Extended Fig. E2). As shown in Fig. 1c and 1d, individual phenalene oligomers with lengths from 2 to 18 units can be easily found, and adsorb planar on Au(111) surface as evidenced by bond-resolved current imaging and nc-AFM imaging (Extended Fig. E3).

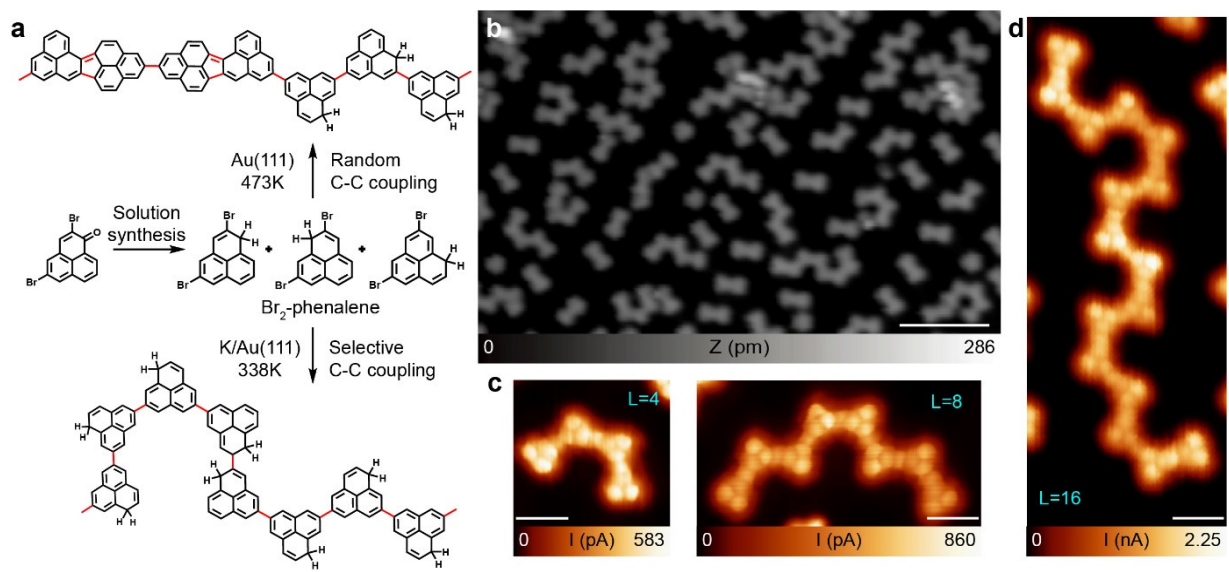

**Fig. 1 | Selective synthesis of individual closed-shell phenalene chains on Au(111). a**, Synthetic scheme of closed-shell phenalene chains via the combination of in-solution and on-surface synthetic methods. **b**, Overview STM image after co-depositing $Br_2$-phenalene and alkali metal potassium on Au(111) held at approximately 300 K and thermal annealing to 338 K (scale bar: 5 nm). **c,d**, Bond-resolved STM images of three representative closed-shell phenalene chains with 4, 8, and 16 close-shell phenalene units ($V_{Bias}$ =5 mV; scale bars: 1 nm).

**Spin-by-spin construction and characterization of [2]triangulene oligomers**

STM tip manipulation allows for the controlled building of finite Heisenberg spin-1/2 chains spin-by-spin within closed-shell phenalene chains. By applying a voltage pulse around 3 V with the STM tip positioned over the sp³ carbon site in a phenalene unit, spins can be generated one-by-one, forming Heisenberg antiferromagnetic coupled spin-1/2 chains. Once the spin chains are constructed, STS differential conductance (dI/dV) spectroscopy is employed to probe the magnetic properties of the chain at the atomic level, which reflects the local density of states under the tip and provides insights into the spin excitation by the second derivative inelastic tunneling process ($dI^2/d^2V$)[57,58].

The chosen [2]triangulene molecule exhibits a spin ground state of S = 1/2, which arises from the sublattice imbalance of the graphene honeycomb lattice (Fig. 2a). This is confirmed by Mean-field Hubbard calculations, revealing a singly occupied molecular orbital and a delocalized spin density distribution, characteristic of a π-radical (Fig. 2a and Extended Fig. E4). The presence of a single spin-1/2 in [2]triangulene is evidenced by the Kondo screening effect showing a zero-bias Kondo peak at the fermi level (Fig. 2d). Kondo resonance imaging reveals the spatial distribution of the unpaired spin, which is delocalized within [2]triangulene unit agreeing with our calculations (Extended Fig. E4).

When [2]triangulene units are covalently linked into a dimer, they form a spin-singlet ground state according to Lieb's theorem (Fig. 2b and Extended Fig. E4). STS dI/dV spectra taken at the dimer exhibit a symmetric conductance step at ±45 meV with respect to the Fermi level, which manifests the excitation gap from the singlet ground state to the excited triplet state as induced by inelastic tunneling electrons (Fig. 2d). This observation evidences the antiferromagnetic coupling between the two unpaired spins with coupling strength of 45 meV, consistent with Mean-field Hubbard calculations.

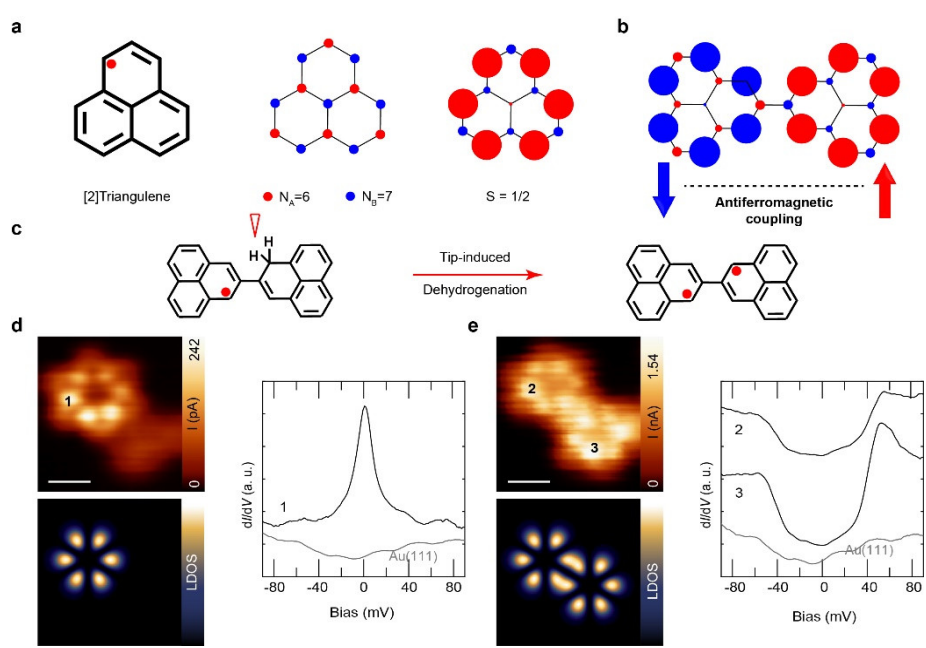

**Fig. 2 | Magnetic properties of [2]triangulene monomer and dimer. a**, Chemical structure of [2]triangulene with a net spin of S = 1/2 due to sublattice imbalance and the corresponding spin-density distribution (red- and blue-filled circles denote intensity of spin-up and spin-down spin density, respectively). **b**, Spin-density distribution of [2]triangulen dimer. The two spins are antiferromagnetically coupled with a coupling strength of 45 meV. **c**, Building of [2]triangulene dimer via atom manipulation. **d**, Constant-height current image ($V_{Bias}$ =5 mV), simulated STM image using SOMO orbital, and dI/dV spectroscopy of [2]triangulene monomer, showing a zero-peak Kondo resonance. **e**, Constant-height current image ($V_{Bias}$=50 mV), simulated STM image and dI/dV spectroscopy of [2]triangulene dimer, exhibiting a U-shape gap due to spin excitation from the singlet ground state to the excited triplet state. Scale bars: 0.5 nm.

Next, [2]triangulene spin chains were constructed one spin at a time within a 9-unit phenalene oligomer, with their spin states traced using STS measurements and elucidated through modeling calculations, as shown in Fig. 3. In finite Heisenberg antiferromagnetic spin-1/2 chains, the evolution of many-body spin states with increasing chain length reveals a complex interplay between quantum confinement and spin interactions, resulting in a rich spectrum of collective spin states. As illustrated in Fig. 3b, the dI/dV spectroscopic features change drastically as spins are added one by one, indicating the presence of many-body collective spin states. For even-numbered chains, a U-shaped gap feature is observed, with the gap size decreasing as the chain length increases. In contrast, odd-numbered chains exhibit a gap feature alongside a zero-bias peak, with the intensity of the zero-bias peak oscillating along the chain. This difference arises from the collective quantum behaviors of spin pairing and highlights the influence of chain length and parity on the magnetic properties of finite Heisenberg spin chains. The presence of zero-bias peaks in odd-numbered chains not in even-numbered chains is due to their different spin ground states (see detailed discussion later). Additionally, we systematically studied spin chains of various lengths, obtaining similar observations (Extended Fig. E5-E7).

We performed exact diagonalization of the Heisenberg spin-1/2 Hamiltonian to elucidate our experimental observations. The calculated spin states are presented in Fig. 3c, which demonstrate a decreasing excitation gap—the energy difference between the ground state and the first excited state—as the chain length increases. For even-numbered chains, the ground state is a spin singlet, with all spins paired and a total system spin of zero. Conversely, odd-numbered chains feature an unpaired spin, resulting in a doublet ground state with a total spin of S=1/2, delocalized along the chain. A perturbation theory, as established by Ternes, was used to simulate the differential spectra of Heisenberg spin-1/2 chains. As shown in Fig. 3d, the simulated spectra closely match the experimental observations, capturing both the gap features and the zero-bias peak. Additionally, we calculated the $\langle S_z \rangle$ weight of the doublet ground state in odd-numbered chains, which exhibited oscillation behavior similar to that observed in experiments (Fig. 3e and f). The strong agreement between experimental results and calculations confirms that the constructed [2]triangulene chains provide a simple and clean platform for studying quantum spin-1/2 antiferromagnets.

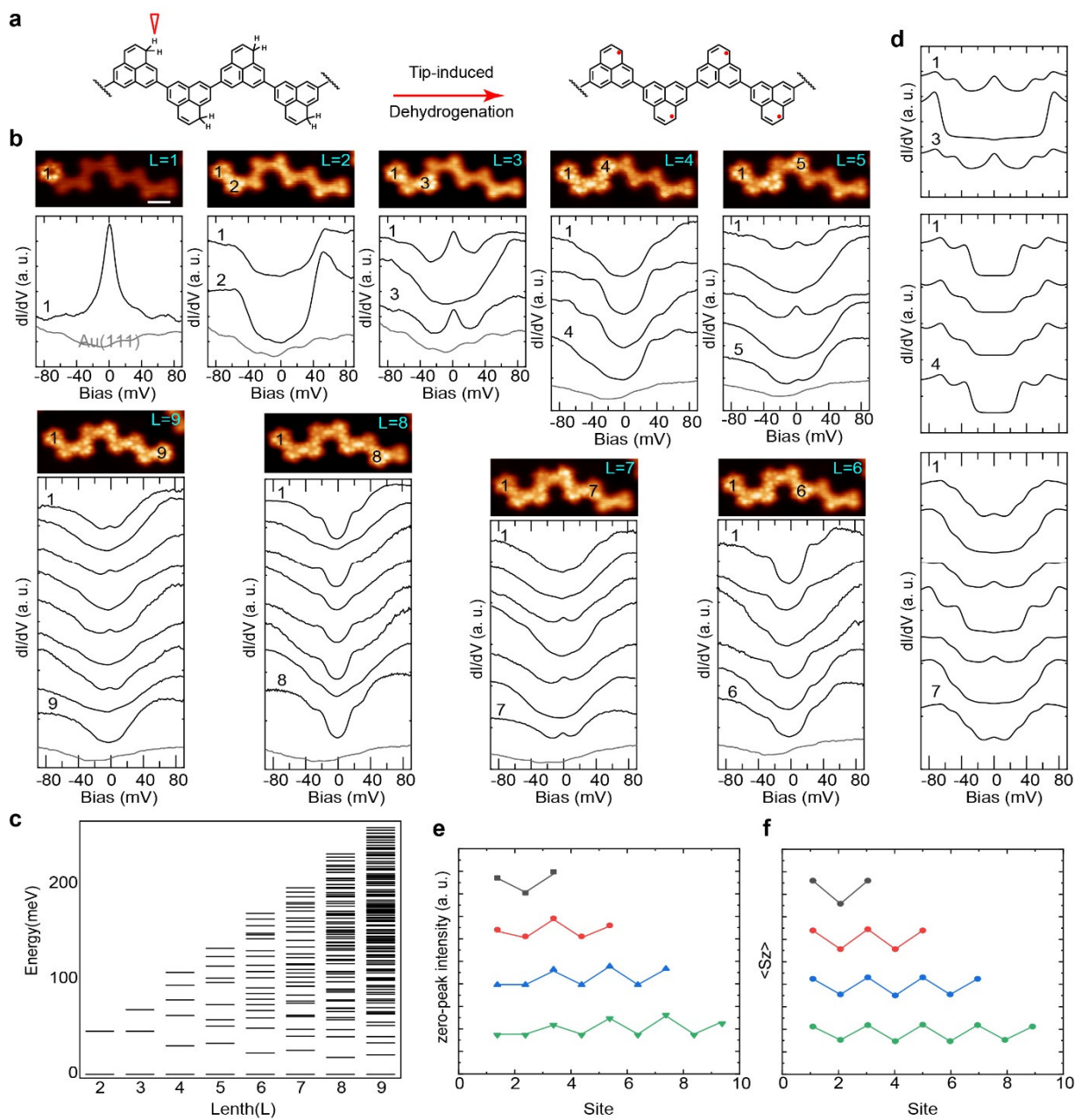

**Fig. 3 | Spin-by-spin building and characterizing Heisenberg antiferromagnetic spin-1/2 chains. a**, Selective formation of finite Heisenberg antiferromagnetic spin chains via step-wise tip-induced dehydrogenation. **b**, Constant-height current images of spin chains with 1-9 [2]triangulene units and the corresponding dI/dV spectra taken on the activated spin sites (scale bar: 1 nm). The current images of odd-numbered chains are taken at $V_{Bias}$ =5 mV, and at $V_{Bias}$ =50 meV for even-numbered chains. **c**, Calculated energy spectrum of spin chains with varying lengths by exact diagonalization of Heisenberg spin chain Hamitonian. **d**, Simulated tunneling spectra using a perturbative approach. **e**, Extracted zero-bias peak intensity of odd-numbered spin chains. **f**, Calculated $\langle S_z \rangle$ weight of the doublet ground state, suggesting the net spin 1/2 is delocalized along the chains.

**Excitation gap evolution of finite Heisenberg antiferromagnetic spin chains**

In finite Heisenberg antiferromagnetic spin chains, the excitation gap exhibits a distinctive evolution as a function of chain length. This gap is influenced by quantum confinement effects and the nature of the spin interactions within the chain. Figures 4a and 4b summarize the excitation gap of even- and odd-numbered spin chains. For shorter chains, the gap is pronounced due to the strong quantum confinement of spinon quasiparticles. As the chain length increases, STS measurements reveal a systematic reduction in the excitation gap. This behavior reflects the decreasing confinement as the chain approaches the infinite limit, where the gap is expected to close entirely, leading to gapless excitations characteristic of an extended one-dimensional quantum antiferromagnet. Furthermore, STS can distinguish between the excitation spectra of odd-numbered and even-numbered chains, highlighting differences in their ground states and the corresponding gap sizes.

We utilize the DMRG method to calculate the excitation gap between the ground state and the first excited state of Heisenberg spin-1/2 chains for lengths up to 200 units, which are difficult to obtain by exact diagonalization because of computation constraints. This approach allows us to accurately capture the behavior of very long Heisenberg spin chains, providing insights into the evolution of the excitation gap as the chain length increases. According to DMRG calculations, the excitation gap in a finite Heisenberg spin-1/2 chain decreases with length following an approximate $1/L$ decay, where $L$ is the length of the chain. This behavior is indicative of how the energy gap between the ground state and the first excited state reduces as the chain lengthens. Specifically, for both even- and odd-numbered spin-1/2 chains, the energy gap decreases as the chain length increases. As shown in Fig. 4c, the obtained experimental excitation gaps nicely fit with DMRG calculations for both even- and odd-numbered chains. For very long chains, this power-law decay causes the excitation gap to approach zero, consistent with the behavior of an infinite Heisenberg chain, which is gapless in the thermodynamic limit (Fig. 4d).

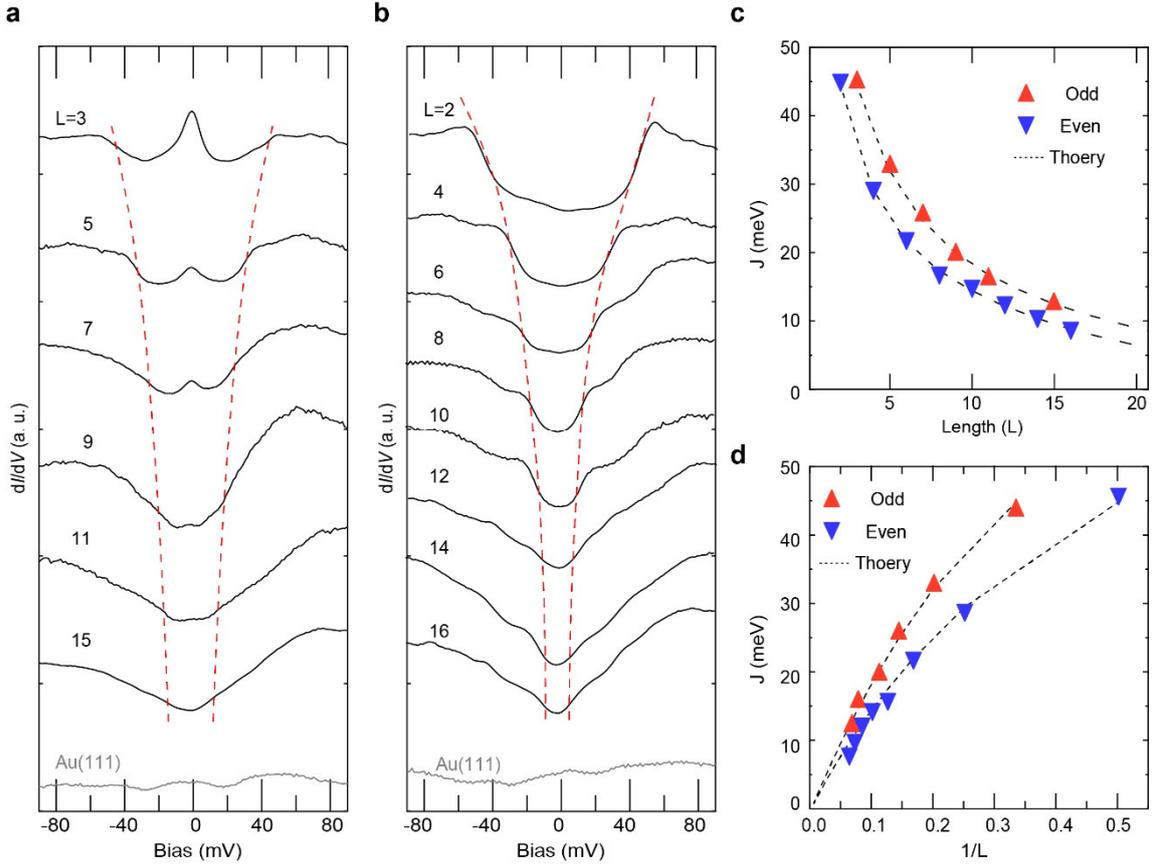

**Fig. 4 | Excitation gap evolution of finite Heisenberg antiferromagnetic spin-1/2 chains. a**, dI/dV spectra taken at the center units of odd-numbered spin chains with varying lengths. Both the gap size and zero-bias peak intensity decrease with increased length. The red dashed line guides the evolution of the excitation gap. **b**, dI/dV spectra taken at the center units of even-numbered spin chains with varying lengths. The red dashed line guides the evolution of the excitation gap. **c,d**, DMRG calculated and measured excitation gap evolution. The gap approximately follows a 1/L power-law decay and reaches to zero for infinite chains in the thermodynamic limit.

### Energy dispersion of confined spinon quasiparticles

As depicted in Fig. 5a, two spinon quasiparticles can be spontaneously generated in a Heisenberg quantum spin-1/2 chain. Inelastic scattering by neutrons or electrons can flip a spin-1/2 site, creating a magnon quasiparticle with spin-1. This magnon would quickly decay into two independent spin-1/2 spinons, which can be envisioned as Néel domain walls dressed by quantum fluctuations. In an infinite spin chain, the energy dispersion of these two spinons fills the region $\omega_l(k) \leq \omega \leq \omega_u(k)$ between the lower and upper energy boundaries[13], where $\omega_l(k) = \frac{\pi}{2} J |\sin K|$ and $\omega_u(k) = \pi J \left|\sin \frac{K}{2}\right|$. As illustrated in Fig. 5b, the energy dispersion reaches its zero energy, at k=0, π and 2π, indicative of gapless excitation, with the lower boundary forming an *m*-shaped feature and the upper boundary displaying a dome-like feature, manifesting a two-particle energy dispersion.

In a finte spin chain, quantum confinement restricts the spinons' motion and quantizes their allowed states, leading to discrete energy levels and a distinctive site-dependent excitation weight as revealed by exact diagonalization of Heisenberg spin-1/2 chain Hamitonian (Fig. 5d). Once the spin states and their spectral weights are computed, a Fourier transform (FFT) of site-dependent excitation weight can be applied to obtain the momentum-resolved energy dispersion of the system. FFT transformed pattern of a 16-unit spin chain reveals an *m*-shaped dispersion of confined spinon quasiparticles with discrete energy and momentum states (Fig. 5d). As the chain length increases, the confinement weakens, leading to a reduction in the energy gap and a closer approach to the continuous dispersion seen in infinite chains (extended Fig. E8).

This behavior has been observed experimentally through the second derivative inelastic tunneling spectroscopy ($dI^2/d^2V$ IETS), which reveals the energy and weight of spin excitations within the system[57-58]. As shown in Fig. 5d, the site-dependent excitation weights of a 16-unit long spin chain provide a clear signature of a pair of spinons confinement in a 1D quantum box. The FFT pattern presented in Fig. 5e reveals an *m*-shaped energy dispersion, consistent with calculations. Similar results were observed in spin chains of varying lengths (Extended Fig. E8), further validating the findings. This detailed analysis through IETS and FFT enables precise characterization of confined spinon quasiparticles, offering an effective strategy to probe quasiparticle excitations in low-dimensional quantum spin systems with combined atomic spatial resolution, high energy resolution, and momentum resolution.

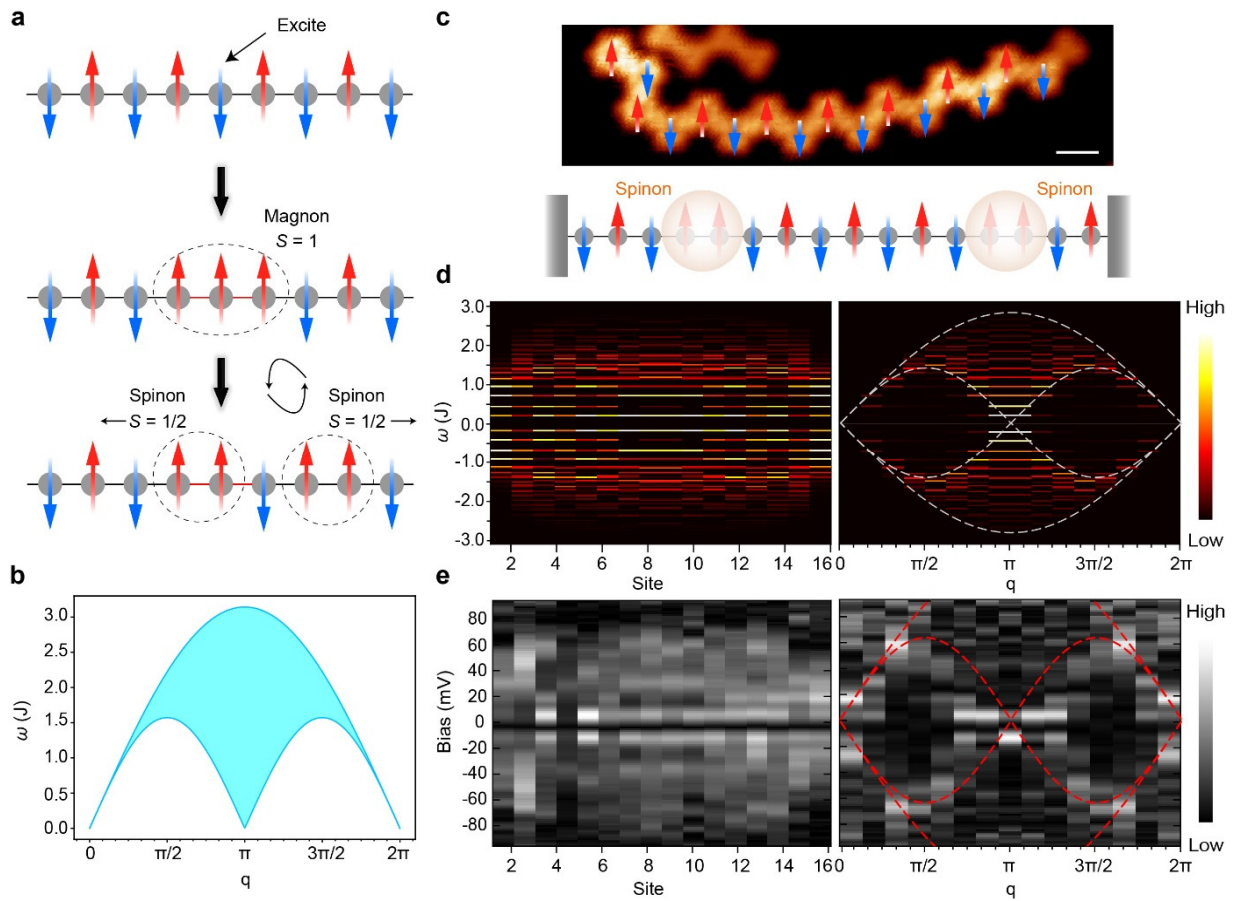

**Fig. 5 | Energy dispersion of confined spinon quasiparticles. a**, Scheme of generating two spinon quasiparticles in an infinite Heisenberg quantum spin-1/2 chain. Spin-flip of one site will generate an intermediate spin-1 magnon quasiparticle, which splits into two spin-1/2 spinon quasiparticles. **b**, Energy dispersion of a pair of spinons in an infinitely long spin chain. **c**, Current image ($V_{Bias}$=40 mV) showing a 16-unit long [2]triangulene chain. Scale bar: 1nm. **d**, Schematic showing the confined two spinon quasiparticles in a quantum box. **e,** Calculated spectra weights of a 16-unit long spin chain, and their FFT image. **f**, Experimental $dI^2/d^2V$ spectra showing the excitation weights of the 16-unit long spin chain in **c**, and their FFT image showing the *m*-shaped energy dispersion. The dashed lines are energy dispersion boundaries of spinons in an infinite spin chain.

**Conclusion**

In conclusion, the atomic-scale characterization of fractional spinon quasiparticles in open-shell triangulene spin chains using STM/STS has provided significant insights into the collective quantum behavior of low-dimensional spin systems. By constructing Heisenberg spin-1/2 chains spin-by-spin and probing their spin states and excitation gaps with atomic resolution, we have directly observed the confinement of spinons and their distinctive *m*-shaped energy dispersion. The ability to visualize and manipulate these quasiparticles in real space represents a major advancement in our understanding of quantum magnetism, particularly in molecular systems. Our findings, consistent with theoretical calculations, not only confirm the presence of fractional spinon excitations but also open up new avenues for exploring their unique properties and potential applications in quantum technologies.

**Methods**

**Synthesis of precursor.** Precursor $Br_2$-phenalene was synthesized from 5-bromo-1*H*-phenalen-1-one by solution methods. The starting material 5-bromo-1*H*-phenalen-1-one was prepared according to the method reported in the literature[59]. The detailed procedure and data are as follows:

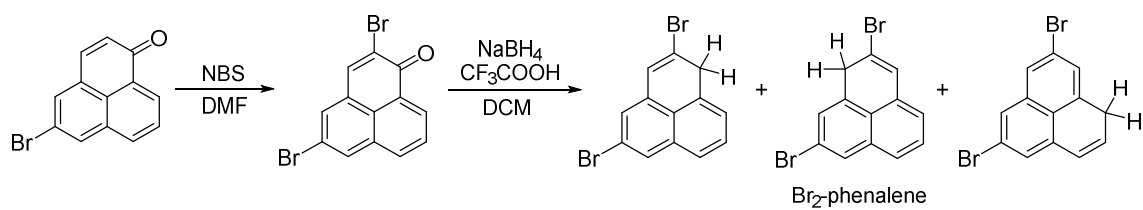

**Synthesis of 2,5-dibromo-1*H*-phenalen-1-one.** To a solution of 5-bromo-1*H*-phenalen-1-one[1] (26 mg, 0.1 mmol) in dry DMF (2 mL) was added NBS (28 mg, 0.16 mmol), and the mixture was reacted at 30 °C for 12 h. The reaction was extracted with ethyl acetate. The organic phase was separated and washed with H₂O, and dried over anhydrous Na$_2$SO$_4$, filtered, the organic phase was combined and concentrated under vacuum, and the residue was purified by chromatography on silica gel (eluent: petroleum ether:CH$_2$Cl$_2$ = 1:1) to afford 2,5-dibromo-1*H*-phenalen-1-one (13 mg, 38%). $^1$H NMR (400 MHz, CDCl$_3$, 25 °C): δ 8.71 (d, *J* =7.36 Hz, 1H), 8.22 (d, *J* =1.32 Hz, 1H), 8.16 (d, *J* =6.2 Hz, 2H), 7.82 (t, *J* =7.32 Hz, 2H); $^{13}$C NMR (150 MHz, CDCl$_3$, 25 °C): δ 178.42, 141.76, 134.65, 133.73, 133.71, 132.27, 132.44, 129.56, 128.68, 128.57, 127.39, 125.37, 120.69; HRMS (EI, TOF): calcd for C$_{13}$H$_6$Br$_2$O$^+$ [M]$^+$: 335,8785, 337.8765, 339.8744, found: 335.8780, 337.8767, 339.8751.

**Synthesis of Br$_2$-phenalene.** NaBH$_4$ (20 mg, 0.5 mmol) and CF$_3$COOH (1.5 mL) were added in a 25 mL reaction tube, and then the mixture was reacted at 40 °C for 20 minutes. After addition of compound 2,5-dibromo-1*H*-phenalen-1-one (34 mg, 0.1 mmol) in CH$_2$Cl$_2$ (4 mL), the mixture was reacted at 40 °C for overnight. The reaction was extracted with CH$_2$Cl$_2$. The organic phase was separated and washed with H$_2$O, and dried over anhydrous Na$_2$SO$_4$, filtered, the organic phase was combined and concentrated under vacuum, and the residue was purified by chromatography on silica gel (eluent: petroleum ether) to afford a mixture of three Br$_2$-phenalene isomers (5 mg, 15%). $^1$H NMR (600 MHz, CDCl$_3$, 25 °C): δ 4.33 (s, 2H), 4.30 (s, 1.9H), 3.96 (s, 0.2H) (The hydrogens on aromatic rings of three Br$_2$-phenalene isomer mixtures were not assigned); $^{13}$C NMR (150 MHz, CDCl$_3$, 25 °C): δ 136.94, 136.28, 134.83, 134.77, 134.18, 133.96, 132.18, 129.94, 129.69, 129.48, 129.05, 128.80, 128.39, 128.24, 128.05, 127.94, 127.68, 127.48, 127.39, 126.48, 126.44, 126.24, 126.16, 125.92, 125.56, 125.41, 125.27, 125.09, 122.89, 122.85, 122.59, 120.42, 41.19, 40.79, 29.85; HRMS (EI, TOF): calcd for C$_{13}$H$_8$Br$_2$$^+$ [M]$^+$: 321.8993, 323.8972, 325.8952; found: 321.8997, 323.8975, 325.8969.

**Sample preparation and STM/nc-AFM measurement.**

A commercially available low-temperature Unisoku Joule-Thomson scanning probe microscope (1.2 K) operated at ultrahigh vacuum(3×10$^{-10}$ mbar) was used for all sample preparation and characterization. The Au(111) single crystals were cleaned using repeated cycles of Ar$^+$ sputtering and subsequent annealing to 950 K to obtain atomically flat terraces. The cleanliness of the crystals was checked by scanning with the STM before molecular deposition. Precursor Br$_2$- phenalene and pure potassium (K, from SAES Getters) were thermally co-deposited on the clean Au(111) surface at room temperature and annealed to 65 ℃ for 5 minutes. Precursor Br$_2$- phenalene was evaporated on the surface from a quartz crucible and the sublimation temperature was approximately 55 °C. Potassium atoms were evaporated through conventional resistance heating of a wire-type K dispenser after complete degassing.

Then, the sample was transferred to a cryogenic scanner at 1.2 K for characterization. Carbon monoxide molecules were dosed onto the cold sample around 9 K ($1\times10^{-8}$ mbar, 1 minute). A lock-in amplifier (531 Hz, 0.1-1 mV modulation) has been used to obtain dI/dV spectra. The STM and STS measurements were taken at 1.2 K, and the data were processed with the WSxM software. The bond-resolved nc-AFM images in Extended Fig. E1 and E3 were acquired with a low-temperature STM (Scienta Omicron POLAR-STM/AFM combined system) operated at approximately 5 K. To achieve ultra-high spatial resolution, CO molecule was picked up from the Au(111) surface to the apex of a tungsten tip. A quartz tuning fork with a resonant frequency of 26 kHz has been used in nc-AFM measurements. The sensor was operated in frequency modulation mode with a constant oscillation amplitude of 0.3 Å. AFM measurements were performed in constant-height mode with *Bias* = 2 mV.

**Density Matrix Renormalization Group calculations.**

We employ the Density Matrix Renormalization Group (DMRG) method to compute the energy gap between the first excited state and the ground state of the Heisenberg chain for lengths ranging from L = 2 to L = 201. The underlying wave-function ansatz of DMRG[60] is Matrix Product States (MPS) [61]. The definition of MPS is

$$|\psi\rangle = \Sigma_{\{\sigma_i\}} Tr(A^{\sigma_1} A^{\sigma_2} \dots A^{\sigma_L})|\sigma_1, \sigma_2, \dots, \sigma_L\rangle$$

where $\sigma_i$ is the local degree of freedom (for s=1/2 spin, it is $|\uparrow\rangle$ or $|\downarrow\rangle$), $A^{\sigma_i}$ is D by D matrix depending on $\sigma_i$. It is known that MPS can efficiently represent the low-energy states of one-dimensional systems. In our calculation, we gradually increase the bond dimension to D = 1000 to ensure the gap is converged with D. For a chain with an even number of sites, the ground state resides in the S = 0 sector, while the first excited state is in the S = 1 sector. In contrast, for a chain with an odd number of sites, both ground state and the first excited state lie in the S = 1/2 sector.

**Exact Diagonalization calculations.**

We also calculate the dynamical spin structure factor using the Exact Diagonalization method. The dynamical spin structure factor in real space is defined as ($|\alpha\rangle$ is the eigenstate of the system with energy $E_\alpha$, $E_g$ is the ground state energy):

$$S^{zz}(j, w) = \Sigma_\alpha |\langle 0|S_j^z|\alpha\rangle|^2 \delta(w - (E_\alpha - E_g))$$

The dynamic spin structure factor in momentum space can be obtained through a Fourier transformation of the spin operator (note that translational symmetry is actually broken in the system with open boundary conditions, the Fourier transformation is employed approximately)

$$S^{zz}(q, w) = \frac{2\pi}{N} \Sigma_\alpha |\langle 0|S_q^z|\alpha\rangle|^2 \delta(w - (E_\alpha - E_g)), S_q^z = \Sigma_j S_j^z e^{-iqj}$$

**Mean-field Hubbard calculations.**

The tight binding (TB) calculation of the STM images was carried out in the C $2p_z$-orbital description by numerically solving the Mean-Field-Hubbard Hamiltonian with nearest-neighbor hopping:

$$\widehat{H}_{MFH} = \sum_{\langle i,j\rangle,\sigma} - t_{ij} c^{\dagger}_{i,\sigma} c_{j,\sigma} + U \sum_{i,\sigma} \langle n_{i,\sigma}\rangle n_{i,\bar{\sigma}} - U \sum_i \langle n_{i,\uparrow}\rangle \langle n_{i,\downarrow}\rangle$$

with $t_{ij}$ is the nearest-neighbor hopping term depending on the bond length between C atoms (For simplicity, we choose $t_{ij}$ = 2.7 eV), and $c^{\dagger}_{i,\sigma}$ and $c_{i,\sigma}$ denoting the spin selective ($\sigma =\uparrow,\downarrow$) creation and annihilation operators on the atomic site $i$ and $j$, U the on-site Hubbard parameter (with $U$ = 3.5 eV used here), $n_{i,\sigma}$ the number operator and $\langle n_{i\sigma}\rangle$ the mean occupation number at site $i$.

**Acknowledgment**

We thank the Ministry of Science and Technology of China (Grants No. 2020YFA0309000), NSFC (Grants No. 92365302, No. 22325203, No. 92265105, 92065201, No. 12074247, No. 12174252, No. 22272050, No. 21925201), the Strategic Priority Research Program of Chinese Academy of Sciences (Grant No. XDB28000000) and the Science and Technology Commission of Shanghai Municipality (Grants No. 2019SHZDZX01, No. 19JC1412701, No. 20QA1405100) for financial support. We also thank the financial support from the Innovation Program for Quantum Science and Technology (Grant No. 2021ZD0302500) and the Shanghai Municipal Science and Technology Qi Ming Xing Project (No. 22QA1403000).

**Author Contributions**

D.-Y.L. and S.W. conceived and supervised the experiments; Y.Z. and Y.J performed the SPM experiments; Y.W. performed the AFM experiments; X.-Y.Z. synthesized the precursor molecules with the supervision of P.-N.L.; X.Q. and M.Q. carried out the theoretical calculations; D.-Y.L. and S.W. wrote the manuscript; All authors discussed the results and commented on the manuscript at all stages.

**Competing Interests**

Authors declare no competing interests.

**Data and Code Availability**

All data and code are available from the corresponding authors upon reasonable request.

**Materials & Correspondence**

Correspondence and requests for materials should be addressed to Mingpu Qin, Peinian Liu, Dengyuan Li, and Shiyong Wang.

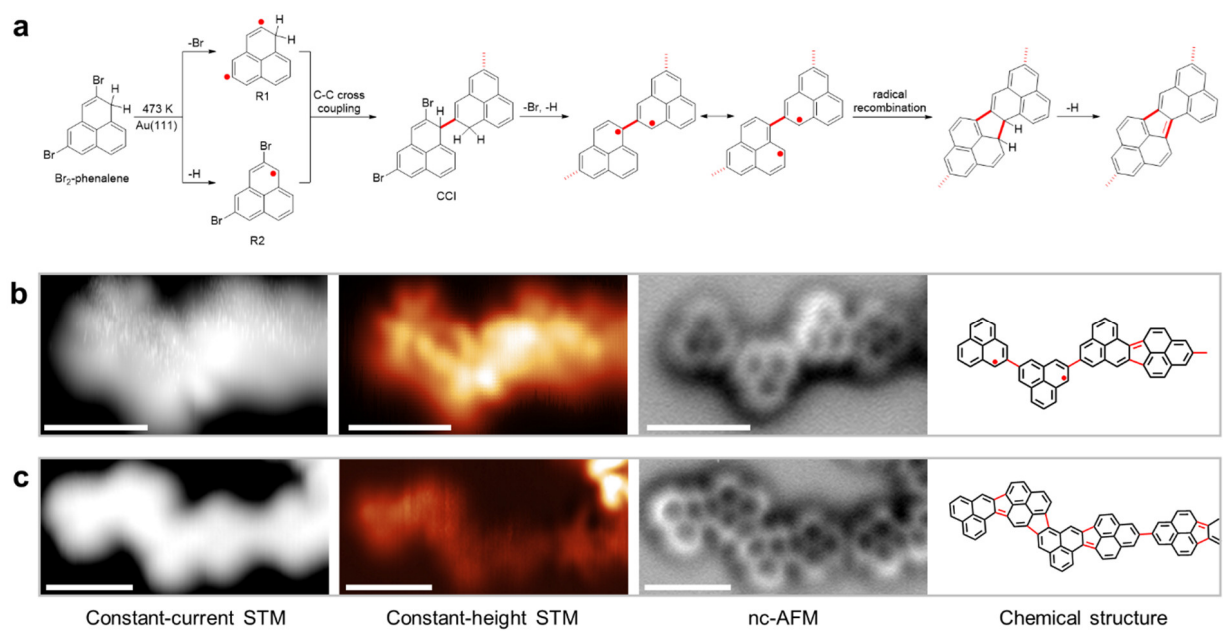

**Extended Fig. E1 | Debrominative and dehydrogenative C-C coupling of Br₂-phenalene on Au(111). a**, proposed reaction pathway for randomly forming five-membered rings on Au(111). After the Au(111) sample containing Br₂-phenalene molecules was annealed at 473 K, Br₂-phenalene may undergo debromination and dehydrogenation to form diverse radical species, such as debrominated radical R1 and dehydrogenated radical R2. Subsequently, these radical species may undergo random radical recombination via C-C homocoupling or cross-coupling, where the cross-coupling of R1 and R2 gives the unexpected coupling intermediate CCI. Finally, CCI undergoes a cascade reaction involving debromination, dehydrogenation, and radical recombination to form unexpected products fused by five-membered rings. **b,c**, Constant-current and -height STM and nc-AFM images of two representative oligomers randomly fused by five-membered rings and the corresponding chemical structures (Scanning parameters of constant-current STM image: $U$ = 100 mV, $I$ = 100 pA). Scale bars: 1 nm.

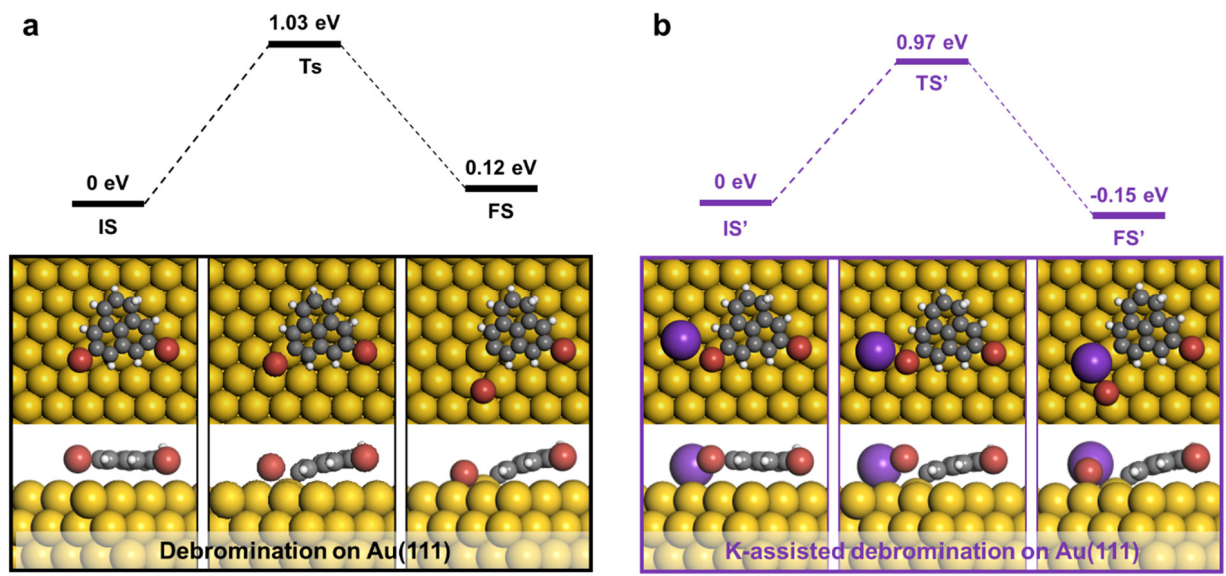

**Extended Fig. E2 | DFT-calculated debromination pathway of Br$_2$-phenalene on Au(111).** **a**, The direct debromination of Br$_2$-phenalene on Au(111). **b**, K-assisted debromination of Br$_2$-phenalene on Au(111). The simulations of the reaction barriers are performed with the climbing image nudged elastic band (CI-NEB) method for finding saddle points and minimum energy paths (http://theory.cm.utexas.edu/vtsttools/neb.html), which is at T = 0 without including entropy and vibrations. The corresponding calculated molecular structure including top and side views of the initial (IS), transition (TS), and intermediate (Int) states of the reactions are shown below the energy diagrams.

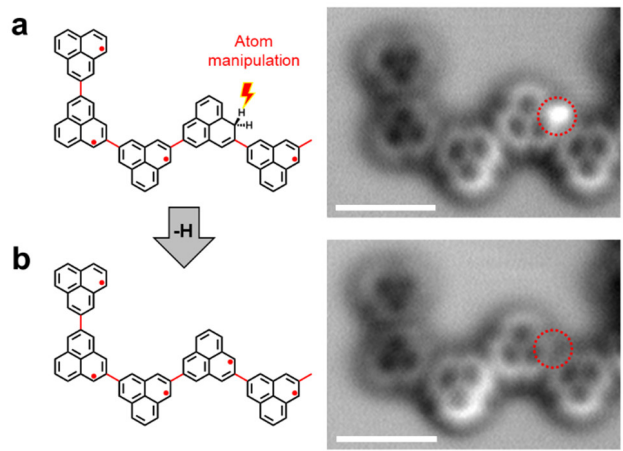

**Extended Fig. E3 | STM tip-assisted dehydrogenation on Au(111). a**, Chemical structure, nc-AFM images of a typical phenalene oligomer with one incomplete dehydrogenation unit. **b**, Chemical structure, nc-AFMimage of a typical spin-1/2 chain formed after atom manipulation. The red dotted circles indicate change in chemical structures before and after atom manipulation. Scale bars: 1 nm.

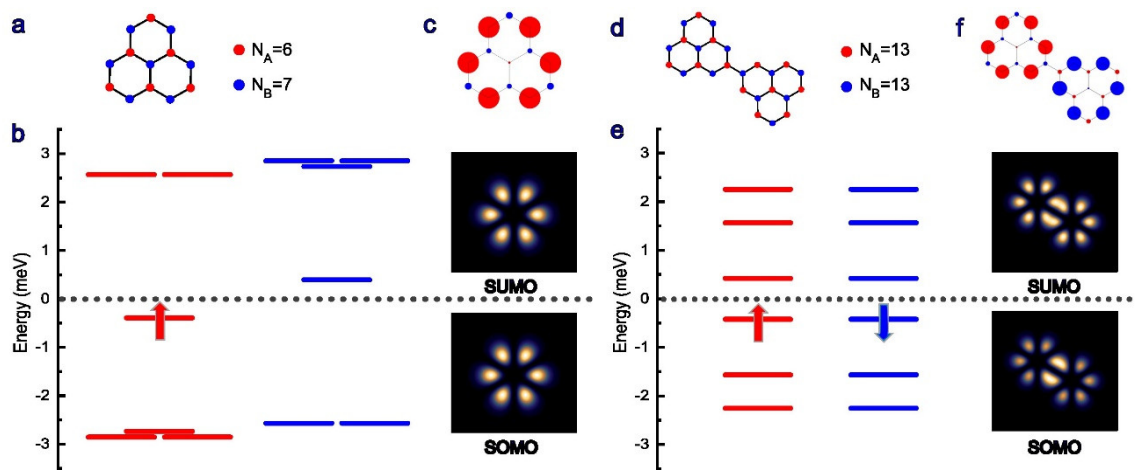

**Extended Fig. E4 | Mean-field Hubbard model calculations. a**, Lattice model of [2]triangulene with sublattice $N_A=6$ and sublattice $N_B=7$, giving a net spin of $S=1/2(N_A-N_B)=1/2$ according to Lieb's theorem. **b**, Calculated energy level, showing a singly occupied/unoccupied orbital (SOMO/SUMO). **c**, From top to bottom: spin density distribution, SUMO and SOMO distribution. Red circles indicate spin-up density and blue spin-down density. **d**, Lattice model of [2]triangulene dimer with sublattice $N_A=13$ and sublattice $N_B=13$, giving a net spin of $S=1/2(N_A-N_B)=0$ according to Lieb's theorem. **e**, Calculated energy level, showing two singly occupied/unoccupied orbitals (SOMOs/SUMOs). **f**, From top to bottom: spin density distribution, SUMO and SOMO distribution. Red circles indicate spin-up density and blue spin-down density.

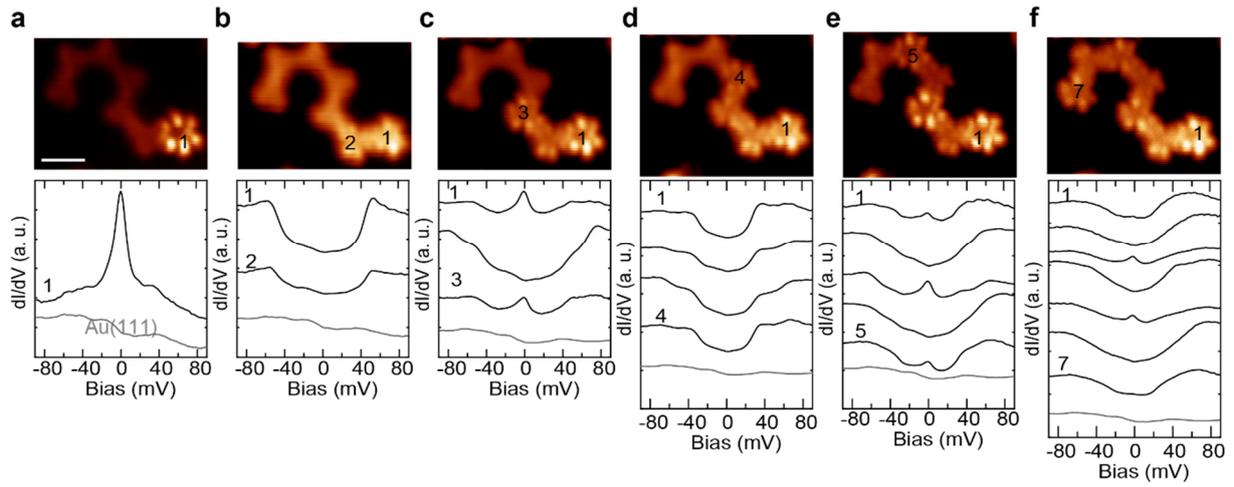

**Extended Fig. E5 | Spin-by-spin building and characterizing Heisenberg antiferromagnetic spin-1/2 chains in a 7-unit phenalene chain. a-f**, Constant-height current images of spin chains with 1, 2, 3, 4, 5, and 7 [2]triangulene units, respectively, and the corresponding dI/dV spectra taken on the activated spin sites (scale bar: 1 nm). The current images in **a**, **c**, **e**, and **f** are taken at 5 meV, in **b**, **d** at 40 mV.

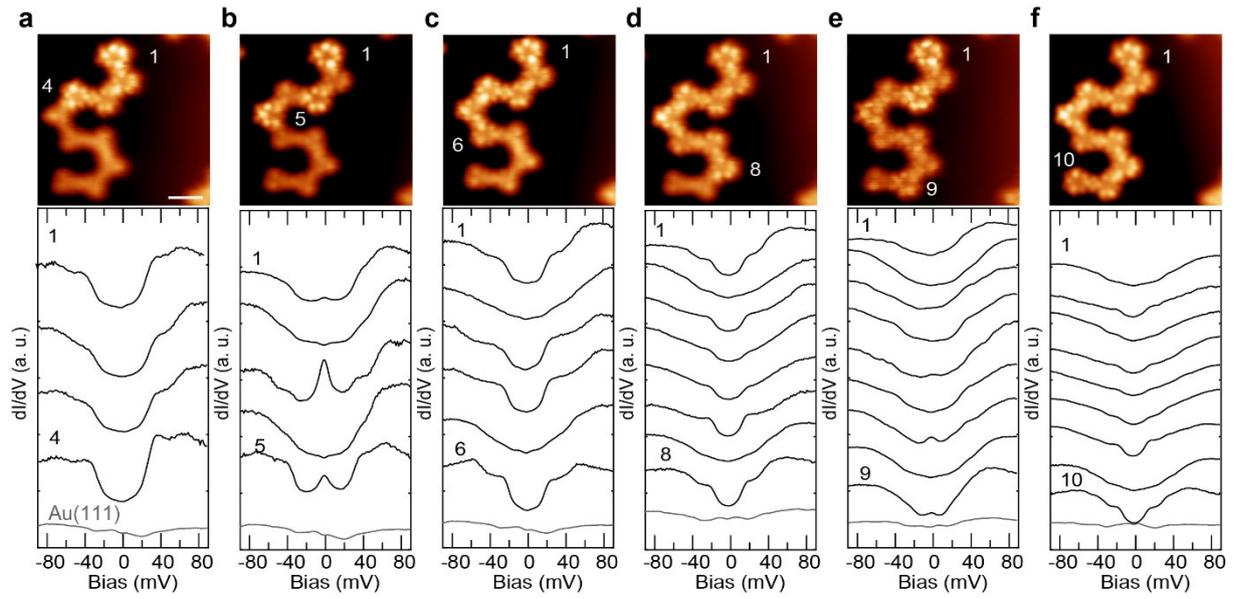

**Extended Fig. E6 | Spin-by-spin building and characterizing Heisenberg antiferromagnetic spin-1/2 chains in a 10-unit phenalene chain.** **a-f**, Constant-height current images of spin chains with 4, 5, 6, 8, 9 and 10 [2]triangulene units, respectively, and the corresponding dI/dV spectra taken on the activated spin sites (scale bar: 1 nm). The current images in **a**, **c**, **e**, and **f** are taken at 5 meV, in **b**, **d** at 40 mV.

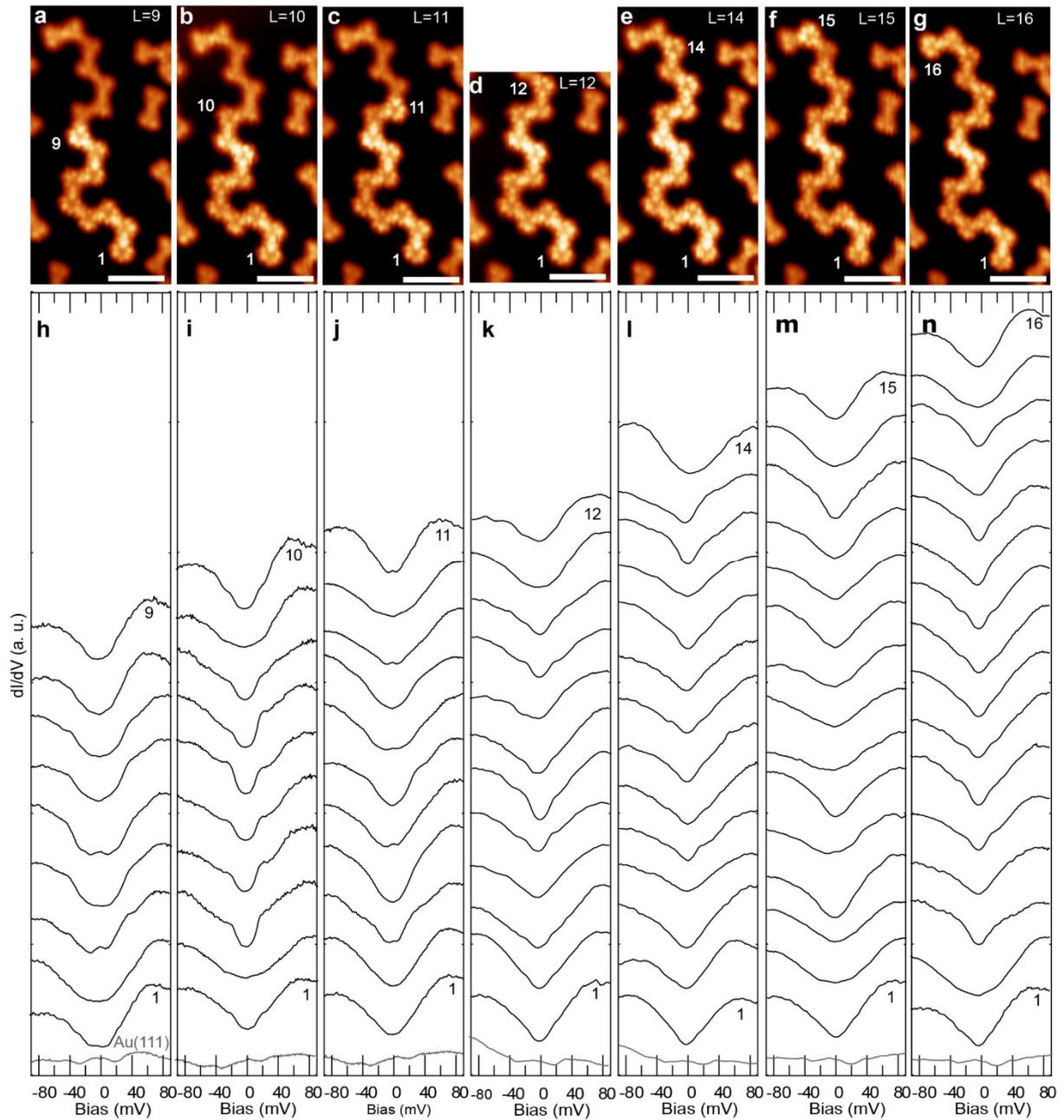

**Extended Fig. E7 | Spin-by-spin building and characterizing Heisenberg antiferromagnetic spin-1/2 chains in a 16-unit phenalene chain.** **a-n**, Constant-height current images of spin chains with 9, 10, 11, 12, 14, 15 and 16 [2]triangulene units, respectively, and the corresponding dI/dV spectra taken on the activated spin sites (scale bar: 2 nm). The current images in odd-numbered chains are taken at 5 meV, in even-numbered chains at 40 mV.

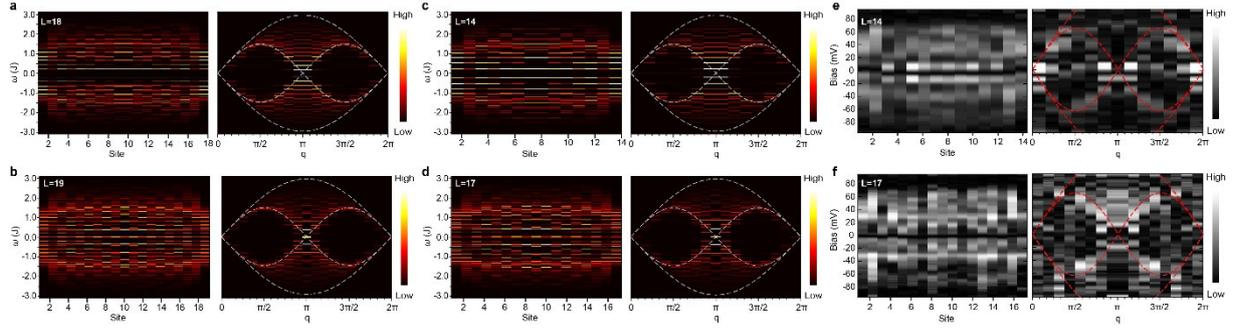

**Extended Fig. E8 | Energy dispersion of confined spinon quasiparticles. a-d,** Calculated spectra weights of a 14-, 17-, 18- and 19-unit long spin chain, and their FFT images. **e-f,** Experimental $dI^2/d^2V$ spectra map showing the excitation weights of a 14- and 17-unit long spin chain, and their FFT images showing the *m*-shaped energy dispersion. The dashed lines are energy dispersion boundaries of spinons in an infinite spin chain. As the chain length increases, the confinement weakens, leading to a reduction in the energy gap and a closer approach to the continuous dispersion seen in infinite chains.